\documentstyle[aaspp4,11pt,flushrt]{article}

\begin{document}

\title{GRB990123: The Case for Saturated Comptonization}

\author{E. P. Liang, A. Crider, M. B\"{o}ttcher, \& I.A. Smith}
\affil{Department of Space Physics and Astronomy, 6100 S. Main,
 	Rice University, Houston, TX 77005-1892}


\newcommand{\Epk}{\rm{E}_{\rm{pk}}}

\abstract{We find that the optical magnitudes of GRB990123 observed by ROTSE 
correlates with the magnitudes extrapolated from the simultaneous 
BATSE spectra, strongly suggesting that the optical, X-ray and 
gamma-ray photons originate from a single source.
We then show that the broadband optical-gamma-ray spectra can be
naturally fit by the saturated Compton model.
We also derive the parameters of the Compton emitting shell from first
principles.
}


\section{Introduction}

The recent simultaneous detection of optical (\cite{aker99}), 
X-ray (\cite{fero99}) and gamma-ray (\cite{kipp99}) 
photons from GRB990123 during the burst provides the first 
broadband multi-wavelength characterization of the burst spectrum 
and evolution.  Here we show that a direct correlation exists between
the time-varying gamma-ray spectral 
shape and the prompt optical emission. This 
combined with the unique signatures of the time-resolved spectra of 
GRB990123 convincingly supports earlier predictions of the saturated 
Comptonization model (\cite{lian97-galactic,lian97-cosmological}).
Contrary to other suggestions (\cite{gala99,sari99}), we find that 
the entire continuum from optical to gamma-rays can be generated 
from a single source of leptons (electrons and pairs). The optical flux
only appears to lag the 
gamma-ray flux due to the high initial Thomson depth of the plasma.  Once 
the plasma has completely
thinned out, the late time afterglow behavior of our model
is the same 
as in standard models based on the Blandford-McKee (1976) solution.
\nocite{blan76}

\section{Data Analysis and Interpretation}

To generate the gamma-ray spectra of GRB990123 during the 
ROTSE observations, we fit the public BATSE data
(SHER/SHERB+DISCSP1; see \cite{pree96}) 
with the Band et al. (1993) \nocite{band93}
GRB function, which is a four-parameter model
that smoothly joins two power laws.  Although the gamma-ray flux
time history initially appears to be uncorrelated with the earliest 
observed optical evolution, a V-band
extrapolation of the Band GRB function
fit to the BATSE data
shows a remarkable correlation with the ROTSE fluxes 
(see Fig. \ref{GRB990123_f1}).  
This suggests that the gamma-ray and optical emission likely arise from
the same source.
We stress that \emph{it is essential to 
consider the time-varying shape of the GRB continuum and not the 
merely the magnitude of the $\gamma$-ray flux.}

Examination of the MER/CONT data during with the first ROTSE time bin,
which includes approximately $10 \times$
as many counts as the SHER/SHERB data, shows an upturn in the lowest
energy channel, inconsistent with the Band GRB function fit by $6.7
\sigma$.
Inclusion of the SD discriminator channel corroborates the MER/CONT upturn
and is inconsistent from the Band GRB function fit by $3.7 \sigma$.
(We also find the discriminator channel data to be in rough agreement with
the peak flux reported by BeppoSAX (\cite{fero99}), assuming a simple
power-law spectra exists between the lowest MER channel and the 
ROTSE observations.)
Fitting the MER+DISCSP1 spectra with a function which allows a low-energy
upturn, such
as the Compton attenuation function (\cite{brai94,brai98}), drastically
improves the $\chi^{2}$ of the fit by 143 without including extra degrees
of 
freedom.  Such upturns have been observed in many previous bursts
(\cite{feni82,brai98,crid99-baas}), for example in GRB970111 using the 
joint BeppoSAX-BATSE spectrum (\cite{crid99-baas}).
As we describe below, however, while the Compton attenuation model can
mimic the observed gamma-ray spectra, it cannot produce the
observed X-rays seen by BeppoSAX.  We turn instead to inverse
Comptonization
spectra, which naturally produces
both the ``terrace'' spectral shape seen by BATSE,
the X-ray flux seen by BeppoSAX, and the optical flux seen by ROTSE.
In Figure 2, we plot the three BATSE spectra deconvolved with inverse
Comptonization spectra (see Section 3),
as well as the simultaneous ROTSE optical measurements.

The popular optically thin synchrotron shock model for the relativistic 
blast wave (e.g. \cite{mesz93,pira93,katz94,tava96})
can only produce the high energy spectral break provided the 
slope below the break is $< -\frac{2}{3}$. 
It cannot produce the additional low energy 
upturn. To maintain the optically thin 
synchrotron shock model, it has been proposed that the optical emission 
comes from a separate component (\cite{sari99}).
However, as we see in Fig. \ref{GRB990123_f1}, a 
separate optical source is unnecessary and artificial given the
correlation between the optical flux and the
gamma-ray extrapolation.  Moreover,
the BATSE data alone in Fig. 2a is enough to 
establish the terrace shape. A separate problem with the synchrotron shock 
model is that it violates the observed spectral evolution seen in many 
bursts (\cite{crid97,pree98-alpha}).
A terrace can in principle be produced also by the Compton
attenuation of 
a power law spectrum by cold intervening material 
(\cite{brai98,brai94,lian94}).
However, the 
extremely high column density ($\rm{{N}_{H} > {10}^{25}~{cm}^{-2}}$)
required, even for sub-solar 
abundances of metals, would have completely absorbed the $< 10$ keV X-rays, 
contrary to the BeppoSAX results.  For GRB990123 the required extinction 
would also be inconsistent with the blue color of the optical source. Thus the 
broadband spectra of Fig. \ref{GRB990123_f2}
are inconsistent with the Compton attenuation model.

The terrace shape continuum, such as that in 
Fig.2a, is a natural consequence and 
unique signature of saturated Comptonization (\cite{rybi79,suny80,feni82})
and was predicted two 
years ago when this model was developed to explain other details of 
GRB spectral evolution (\cite{lian97-galactic,lian97-cosmological}).
In this picture the GRB phase of a burst 
corresponds to the time when the relativistically expanding shell is still very 
dense and Thomson thick ($\rm{\tau_{T} > 1}$).
\emph{The X- and gamma-rays are produced by 
multiple Compton upscattering of self-emitted synchrotron and 
bremsstrahlung soft photons} (peaking below the IR;
\cite{lian97-cosmological}).  When the Thomson 
depth is very high a Wien peak emerges producing a high-energy spectral 
break (\cite{rybi79}).
At low energies, the spectrum turns up with a power-law
of slope $\leq -1$, 
producing the characteristic terrace shape (Fig. 2a \& 3a).  The low energy
upturn
usually occurs approximately one order of magnitude below the
high energy 
break (\cite{rybi79,suny80,lian84}).
This low energy upturn is caused by the competition between 
spatial and energy diffusion in a medium with \emph{volume} injection of soft 
photons.  
In the Compton cooling model
(\cite{lian97-galactic}), the photon diffusion
time is assumed to be \emph{shorter} than the
observable burst evolution timescales.  The 
typical ``hard-to-soft'' spectral evolution
and pulse broadening with decreasing photon
energy are then interpreted as due to the 
Compton cooling plus optical thinning of the 
emitting plasma.  The above limit on the
photon diffusion time puts constraints on the
shell thickness and density.  Details will
be discussed elsewhere.

\section{Monte Carlo Spectral Modeling}

We have performed a systematic modeling of the three GRB990123 
spectra using our Monte Carlo Compton code for hybrid thermal-nonthermal 
plasmas (\cite{bott98}).
A thin uniform shell of thermal plus nonthermal leptons Compton
upscatter self-emitted synchrotron and bremsstrahlung photons.  The 
output is then Lorentz boosted and cosmologically redshifted to the observer 
frame to match the data.  Hence the shell parameters are dependent on the 
bulk Lorentz factor which must be constrained from other physical 
self-consistency considerations (e.g. pulse rise times; 
\cite{lian97-cosmological}).
For a given $\Gamma$ the Comptonized spectrum
is then specified by the (comoving frame) magnetic field B, 
Thomson depth $\tau_{\rm{T}}$,
thermal lepton temperature T, nonthermal lepton fraction $\xi$
and nonthermal power law index p.  We have used an upper Lorentz 
factor cutoff of ${10}^{6}$
in the lepton distribution, which is adequate for modeling spectra 
up through the COMPTEL energy range. 

Because of the large number of model parameters,
the currently available 
spectral data is not very constraining on the model parameter space.
Even if 
we had more data points (e.g. from BeppoSAX, OSSE, and COMPTEL) 
computational limitations prohibit us from directly
searching for the best fit model spectrum using chi-square
minimization techniques.
Hence the model fits presented here in Fig.2 
are only meant to demonstrate a proof of principle, showing the 
resulting terrace shape and its evolution.
Parameters of the sample model are
listed in the caption of Figure 2.  Even 
with such crude fits we can see evidence of Thomson thinning
and steepening of the nonthermal lepton index.

In Fig.3 we show examples of different Monte Carlo Compton spectra 
to illustrate how the spectral shape varies with the different 
input parameters.  We see that the prominence of the terrace is
primarily correlated with the Thomson depth.    

\section{Compton Shell Parameters}

We have deconvolved the time-resolved BATSE spectra for the whole burst 
using the Band et al. GRB function.
The BATSE spectral evolution data allows us to derive the $\Epk$-fluence
decay constant (\cite{lian96,crid99-catalog}) $\Phi_{0}$ of each 
pulse.  We find apparent $\Phi_{0}
\sim 150-300 {\rm{cm}}^{-2}$ for the pulses containing 
the three optical intervals (the intrinsic $\Phi_{0}$ could be
lower due to gravitational lensing).  From this plus the pulse 
rise times we can constrain from first principles the Compton ejecta
shell parameters (\cite{lian97-cosmological}),
including $\Gamma$, the shell radius, thickness, density,
total leptonic mass $\rm{{M}_{e}}$
and total leptonic 
energies.
These model parameters are listed in Table 1.
Using these parameters we can 
estimate the transition time from the free-expansion (internal shock)
phase to 
the blast wave (external shock) phase which is nominally identified with the 
``afterglow'' power law, as a function of the ISM/CBM density and ejecta 
proton loading. We see that
the predicted transition time to the afterglow phase (cf. Table 1)
is indeed consistent with the observed transition time of a few hundred
seconds (\cite{gala99}).

\acknowledgements
This work was supported by NASA (NAG 5-3824).  A.C. thanks MSFC for 
his NASA GSRP fellowship.

\clearpage
\figcaption [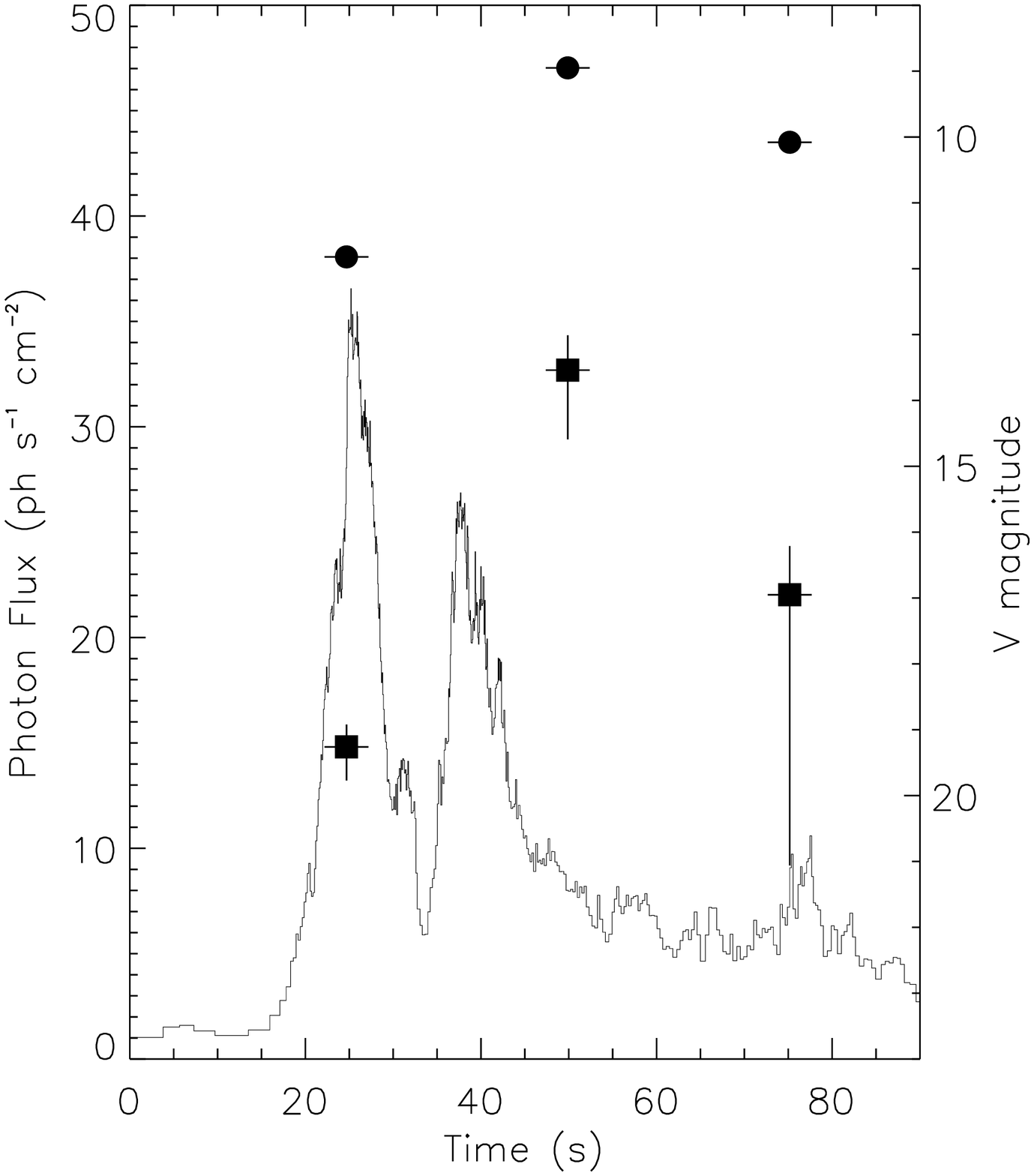] {The gamma-ray (24-1800 keV)
and optical prompt emission of GRB990123.
The gamma-ray photon flux is calculated from MER data fit using a Band et al.
(1993) 
GRB function with $\beta$ fixed to -3 (as found by OSSE).  Extrapolating a Band 
GRB function fit to the SHER/SHERB data during each ROTSE time bin 
gives V magnitudes (plotted as squares). This shows a similar evolution to 
the ROTSE observations (plotted as circles). There are minor differences in 
the time intervals used due to binning constraints.
\label{GRB990123_f1}}

\figcaption [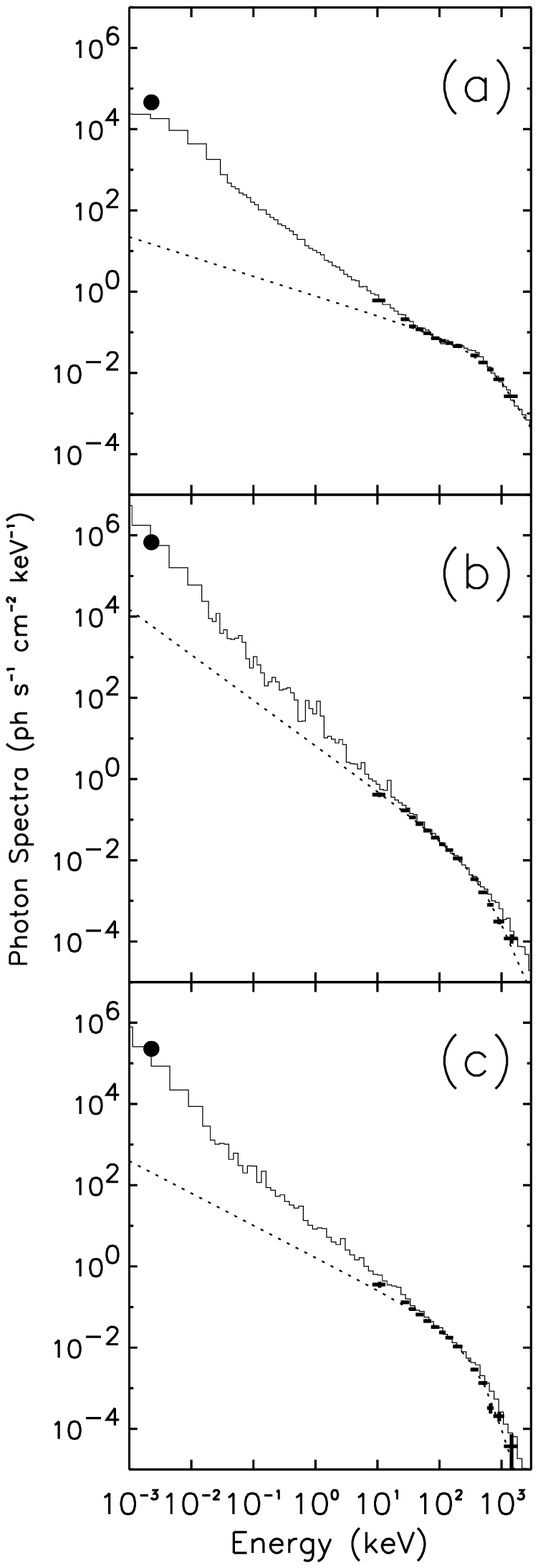] {The BATSE spectra of GRB990123 coincident with the
first three 
ROTSE observations. The dashed curves show the extrapolations of the 
Band (1993) GRB function fits to the gamma-ray data. 
The histograms show our 
Monte Carlo inverse Comptonization model spectra. The parameters used in each 
simulation are: (a) $\rm{\tau_{T} = 20, kT~\Gamma / (1+z) = 80~keV, \xi = 3\%,
B={10}^{2}~G, p = 3}$;  
(b) same parameters except $\rm{\tau_{T}}$ = 6 and p = 6;
(c) same parameters except $\rm{\tau_{T}}$ = 8.4 and p = 6.  
\label{GRB990123_f2}}

\figcaption [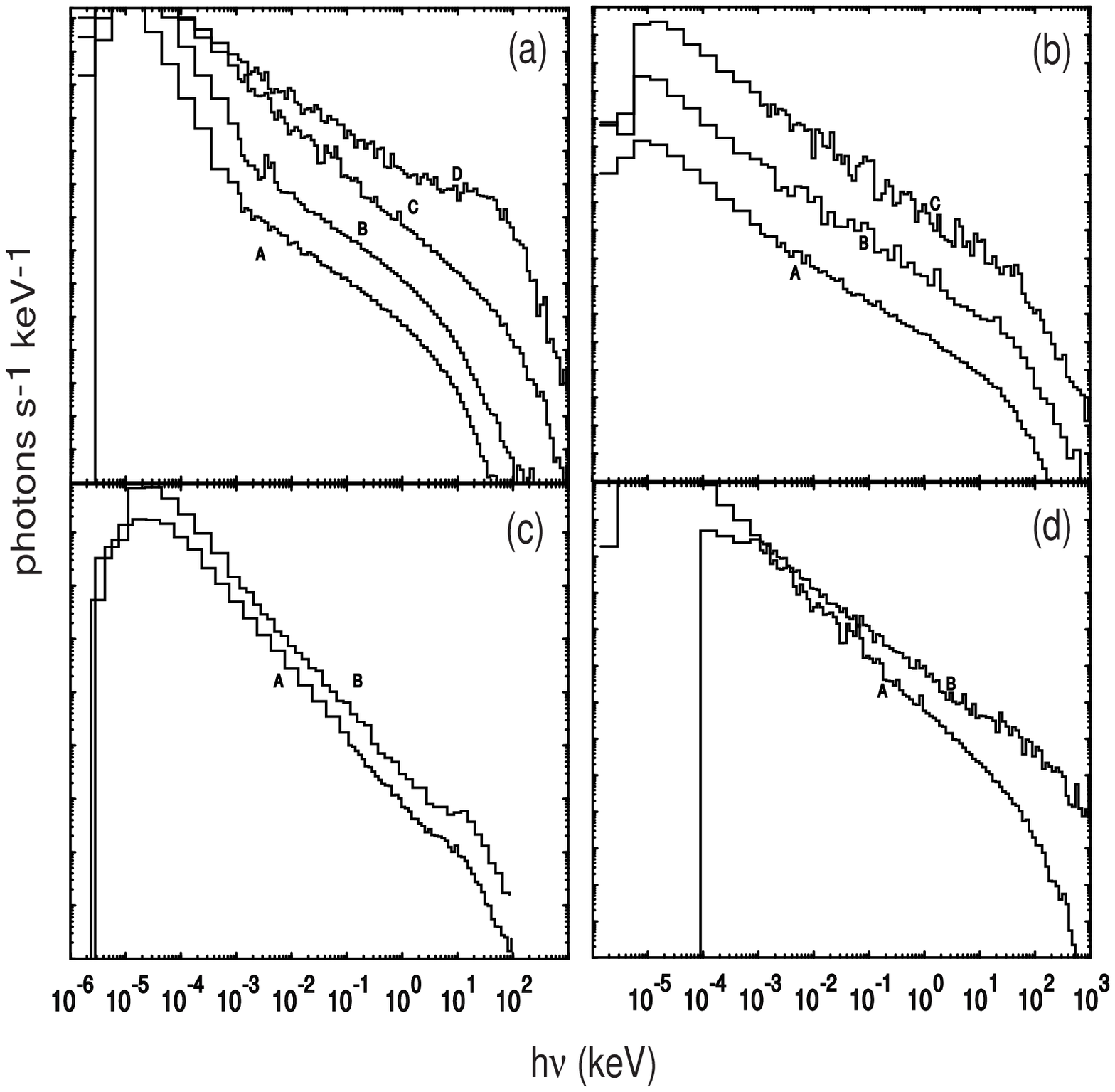] {Compton spectra from a hybrid
thermal-nonthermal plasma code (a) varying Thomson depth
$\rm{ \tau_{T} = [10^{-5}(A),1(B),6(C),20(D)]}$
with kT=5 keV, $\xi$=10\%, B=$10^3$ G, and p=6. 
Note that the uncomptonized spectrum (curve A) is a superposition
of synchrotron and bremsstrahlung source functions;
(b) varying nonthermal fraction  
$\xi$ = [3\% (A), 5\% (B), 10\% (C)]
with kT=5 keV, $\tau_T$=8.4, B=$10^3$ G,and  p=6. 
Note that the spectral break moves to higher energy as 
$\xi$ is increased;
(c) varying thermal temperature kT=[1 keV (A), 3 keV (B)]
with $\tau_T$=16, B=$10^2$ G, and p=3.
Note that both the spectral break moves to higher energy and the spectrum
gets harder as kT is increased;
(d) varying nonthermal lepton index
p=[6 (A), 2 (B)]
with $\tau_{T}$=6, kT=5 keV, B=$10^3$ G, and $\xi$=10\%.}

\clearpage
\plotone{f1.ps}
\clearpage
\plotfiddle{f2.ps}{1in}{0}{100}{100}{-325}{-625}
\clearpage
\plotone{f3.eps}

\clearpage

\newcommand{\dist}{\rm{d}_{9}}
\newcommand{\phit}{\Phi_{200}}
\newcommand{\taut}{\tau_{20}}
\newcommand{\delt}{\Delta \rm{T}_{10}}

\begin{table}
\begin{tabular}{rlcl}

Radius  & R &   $=$ &   $\rm{
                        8.5 \times 10^{16}~cm~d_{9}~\Phi_{200}^
                        {1/2}~\tau_{20}^{-1/2}
                        }$ \\

Bulk Lorentz factor & $\Gamma$ & $\geq$ & $ 376~\dist^{1/2}~\phit^{1/4}~\taut
                        ^{-1/4}~\delt^{-1/2} $ \\

Total lepton number & $\rm{N_{e}}$ & $=$ & $ 2.7 \times 10^{60}~\dist^{2}~
                        \phit~\Omega_{4 \pi} $ \\
        
Total lepton Mass & $\rm{M_{e}}$ & $=$ & $ 2.5 \times 10^{33}~\rm{gm}~
                        \dist^{2}~\phit~\Omega_{4 \pi} $ \\

Total shell mass & M & $=$ & $\rm{
                        M_{e}~f_{p} \equiv
                        M_{e} (1 + 1836 \frac{n_p}{n_e}) \leq 1837 M_{e}}$ \\

Total bulk kinetic energy & $\rm{M~\Gamma~c^{2}}$ & $\geq$ &
                        $\rm{8.5 \times 10^{56}~ergs~\dist^{5/2}~
                        \phit^{5/4}~\taut^{-1/4}~\Omega_{4 \pi}~f_{p}~
                        \delt^{-1/2} }$ \\

Blastwave transition radius & $\rm{R_{BW}}$ & $\equiv$ &
                        $\rm{ (\frac{3~M}{4 \pi~\Omega_{4 \pi}~m_{H}~n_{ISM}})
                        ^{1/3} = 
                        7.1 \times 10^{18}~cm~\dist^{2/3}~
                        \phit^{1/3}~f_{p}^{1/3}~n_{ISM}^{-1/3}}$ \\
 
Blastwave transition time & $\rm{T_{BW}}$ & $\equiv$ &
                        $\rm{ \frac{R_{BW}}{2 \Gamma^{2} c} \leq
                         837~sec~\dist^{-1/3}~\phit^{-1/6}~
                         \taut^{1/2}~f_{p}^{1/3}~\delt~n_{ISM}^{-1/3}}$ \\

\end{tabular}

\caption{Parameters of the GRB ejecta shell based on the saturated Compton
model (scaled to the spectral parameters of the first ROTSE
interval). Here $\dist \times
9~\rm{Gpc}$ is the distance to the burster, $\phit \equiv \Phi_{0}/200$,
$\taut \equiv \tau_{\rm{T}}/20$, $\rm{T}_{10} \times 10~\rm{sec}$
is the pulse rise time in the detector frame,
$\Omega_{4 \pi}$ is the shell angular filling
factor divided by $4 \pi$, $\rm{n_p / n_e}$ is the ratio of ejecta protons to
leptons, and $\rm{n_{ISM}}$ is the ISM density in $\rm{cm^{-3}}$.}

\end{table}

\end{document}